\documentclass[a4paper,fleqn,usenatbib]{mnras}
  
\usepackage[T1]{fontenc}
\usepackage{ae,aecompl}
\usepackage{lineno}
\usepackage{threeparttable}

\usepackage{epstopdf}
\usepackage{graphicx}
\usepackage{lineno}
\usepackage{amssymb}	% Extra maths symbols
\usepackage{amsmath}

\newcommand{\rmd}{{\rm d}}

\title[Successive Injection of Opposite Magnetic Helicity in AR 12257]{ Successive Injection of Opposite Magnetic Helicity: Evidence for Active Regions without Coronal Mass Ejections }

\author[P.~Vemareddy]{
	P.~Vemareddy$^{1}$\thanks{E-mail: vemareddy@iiap.res.in}
	\\
	$^{1}$Indian Institute of Astrophysics, Sarjapur road, II Block, Koramangala, Bengaluru-560 034, India
}
\date{Accepted XXX. Received YYY; in original form July 6, 2021}

\pubyear{2021}

\begin{document}
	\label{firstpage}
	\pagerange{\pageref{firstpage}--\pageref{lastpage}}
	\maketitle
	
%\author[0000-0003-4433-8823]{P.~Vemareddy}

%%%%%%%%%%%%%%%%%%%%%%%%%%%%%%%%%%%%%%%%%%%%%%%%%%%%%
%% Abstract %
%%%%%%%%%%%%%%%%%%%%%%%%%%%%%%%%%%%%%%%%%%%%%%%%%%%%%
\begin{abstract}
Magnetic helicity (MH) is a measure of twist and shear of magnetic field. MH is injected in the active region (AR) corona through photospheric footpoint motions causing twisted and sheared magnetic fields. From the conservation property of the helicity, it was conjectured that an already twisted flux rope (FR) with continuous injection of MH inevitably erupts to remove the excess accumulated coronal helicity. Therefore, understanding the nature and evolution of the photospheric helicity flux transfer is crucial to reveal the intensity of the flare/CME activity. Using the time-sequence vector-magnetograms of \textit{Helioseismic Magnetic Imager}, we study the evolution of MH injection in emerging AR 12257. The photospheric flux motions in this AR inject positive helicity in the first 2.5 days followed by negative helicity later. This successive injection of opposite helicity is consistent with the sign of mean force-free twist parameter ($\alpha_{av}$), orientation of magnetic-tongues. Also, the extrapolated AR magnetic structure exhibits transformation of global-shear without a twisted FR in the core of the AR. No CMEs are launched from this AR but C-class flaring activity is observed predominantly in the second half of the evolution period. The ARs with sign reversal of the MH injection are not favorable to twisted FR formation with excess coronal helicity and therefore are important to identify CME-less ARs readily. A possible scenario in these ARs is that when one sign of helicity flux is replaced by opposite sign, the magnetic field of different connectivity with opposite shear undergoes reconnection at different scales giving rise to both intermittent flares and enhanced coronal heating.
\end{abstract}

\begin{keywords}
Sun: flares --- Sun: coronal mass ejection --- Sun: magnetic fields --- Sun: magnetic reconnection --- Sun: magnetic helicity
\end{keywords}

%%%%%%%%%%%%%%%%%%%%%%%%%%%%%%%%%%%%%%%%%%%%%%%%%%%%%%%%%
%% 1. Introduction %
%%%%%%%%%%%%%%%%%%%%%%%%%%%%%%%%%%%%%%%%%%%%%%%%%%%%%%%%%
\section{Introduction}
\label{Intro}

%\linenumbers

%   {\S}{\bf --- Activity, photospheric motions} \\
The large-scale activity of the sun, i.e. flares and coronal mass ejections (CMEs), is believed to occur by release of the magnetic energy which was pre-stored in the slowly evolving magnetic field of the active regions (ARs). The evolution of the photospheric magnetic field is driven by the slow motions of the footpoints. These motions are dominated by vertical plasma motions in the early phase of the AR emergence and later on dominated by horizontal motions. These motions include in various ways shear, twist, converging, and divergent motions. They induce the foot point motions of coronal flux tubes, which imply a transfer of magnetic energy, shear and twist in the coronal magnetic field \citep[e.g.][]{Kusano2002,priest2002,tian2008a}.

The amount of twist and shear of the magnetic field is quantified by a physical parameter called magnetic helicity \citep{Mofatt1978_book}. The magnetic helicity is conserved in an ideal MHD process and changes very slowly in a resistive process \citep{Taylor1974, Berger1984_Hcons, Pariat2015_TestMagHelConsr}.  
%Because of gauge invariance, in the solar context, (The gauge invariance is rather linked to the needed definition of relative H: so to introduce) time derivative of 
Using this conservation property, the photospheric flux of relative magnetic helicity was employed to estimate the accumulation of the coronal helicity budget by photospheric foot point motions \citep{Chae2001_ObsDetMagHel, Demoulin2002_Hbudget, pariat2005, tian2008b, Liuyang2012_HelEner, Vemareddy2015_HelEne}. Early studies focused on the long term injection by differential rotation \citep{DeVore2000_HdiffRot, Demoulin2002_MagHelShMot} as well as on the impulsive large variations of the helicity injection during M and X-class flares \citep{MoonYJ2002_ImpulsiveVar}. A statistical study was also carried out to relate the variation of the helicity injection rate and its accumulation with the flare productivity \citep{Labonte2007_SurveyMagHel,Park2010_ProdFlareMagHel}. 

Present understanding of eruptive activity links the origin and existence of the twisted flux rope without which large scale CME eruption beyond 10\,R$_\odot$ is not possible \citep{Vourlidas2013}. Recent observations of the ARs reveal flaring activity without the associated CMEs, which require a different physical mechanism than CME producing ARs. From this perspective, the connection of the helicity flux evolution with a distinct activity had advanced our understanding recently. Using time-sequence high-resolution vector magnetic field observations of the ARs, \citep{Vemareddy2015_HelEne} first discussed the global time-profile of the helicity flux injection and its relevance to CMEs. In this report, he presented three kinds of emerging ARs with helicity flux injection positive/negative and a changing sign over the time-evolution. The ARs with positive/negative helicity flux injection were found to launch CMEs at some point in time, whereas the AR with sign-changing helicity flux injection launched no-CMEs except C-class flares. Further such studies reported that the ARs with a predominant sign of helicity injection are the source regions of the CMEs (e.g., \citealt{Vemareddy2017_homologus,vemareddy2019_VeryFast,DhakalS2020}). These studies inferred that the magnetic flux rope is built up by the line-tied photospheric motions which inject the magnetic helicity continuously. 

From the basic conservation property of the helicity, \citet{zhangmei2005} proposed that an already formed twisted flux rope (twisted flux) with continuous injection of magnetic helicity inevitably erupts in order to remove the excess accumulated helicity. Given this idea implies to the existence of helicity upper bound in the AR which is still at a conjecture level. Further this Low's proposal of helicity upper bound was investigated in a theoretical study by \citet{Zhangmei2008_HelBound} which suggested that a coronal magnetic field may erupt into a CME when the applicable helicity bound falls below the already accumulated helicity as the result of a slowly changing boundary condition. Their calculations also showed that a monotonic accumulation of magnetic helicity can lead to the formation of a magnetic flux rope applicable to kink instability. 

Towards the above Low's conjecture of helicity upper bound, numerical simulations have also been performed. \citet{Zuccarello2018_Thres_MagHel} used MHD simulations to evaluate the accumulation of relative magnetic helicity under the different boundary flows. Their study suggested the existence of a threshold in a quantity defined by the ratio between the non-potential magnetic helicity and the total relative magnetic helicity, which remained same at the onset of the eruptions in all simulation runs. This ratio was also indicative to be eruptive potential in the AR magnetic field constructed from non-linear force-free field model based on observed photospheric vector magnetic field \citep{Moraitis2019_MagHel_erup,Thalmann2019_MagHel_ARs}.  

In the above context, therefore, the nature of the helicity injection with a predominant sign is the key to reach the state of the excess coronal helicity accumulation and then the CME eruption from the AR. On the other hand, very few ARs are seen with helicity injection whose dominant sign is changing in time \citep{Yamamoto2009, LiuYang2014_MagHelEmerARs, Vemareddy2015_HelEne}. In that case, the helicity accumulation in the first part of evolution is possibly canceled by the later part. In an emerging AR 11928, the helicity flux was found to change sign from positive to negative over its evolution \citep{Vemareddy2015_HelEne, Vemareddy2017_SucHelInj}. CMEs were not launched from this AR except the C-class flaring activity. To the best of our knowledge, this was the first study citing a counter example to the eruptive ARs with predominant sign helicity injection. Such reports provide observational proof of conservation property of the helicity and will shed more light on the threshold levels of helicity budget in the ARs. In this letter, we report another AR case of changing sign of helicity injection over time and the scenario of coronal magnetic field evolution. Observational data used are briefly described in Section~\ref{ObsData}, results are presented in Section~\ref{Res} and a summarized discussion is given in Section~\ref{SumDisc}.

\begin{figure}  %---------------------------------------------------------------------
    \centering
	\includegraphics[width=.5\textwidth,clip=]{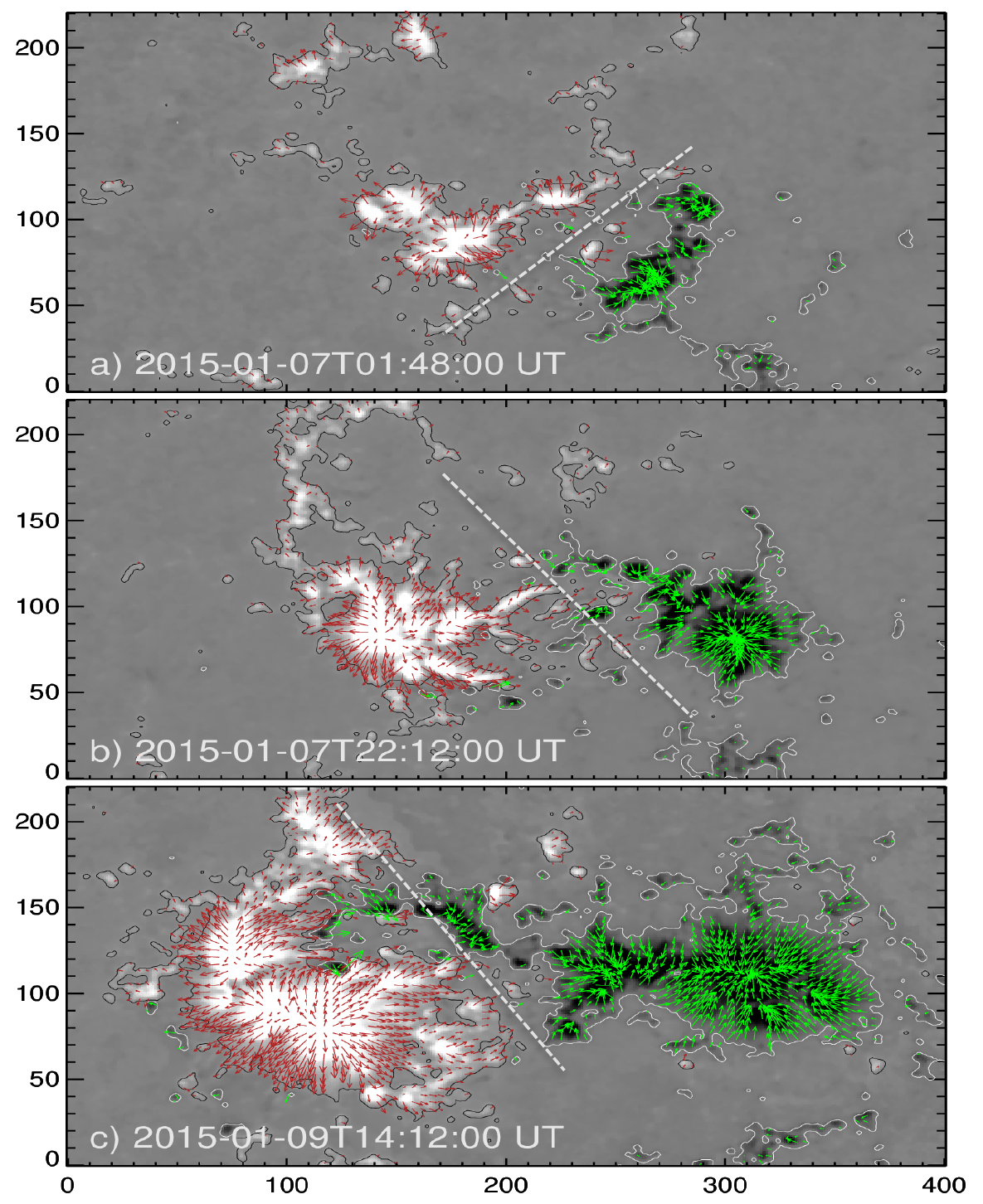}
    \caption{HMI vector magnetic field observations of AR 12257 at different times during its emergence. The background image is the normal component of the magnetic field with contours at $\pm$120 G. Green/red arrows refer to horizontal fields, ($B_x$, $B_y$), with their length being proportional to the magnitude $B_h=\sqrt{B_x^2+B_y^2}$. White-dashed line indicates the orientation of the magnetic tongue, which is same as the PIL. The axes units are in pixels of 0.5 arcsec. }
    \label{Fig_VMgram}
\end{figure}

%%%%%%%%%%%%%%%%%%%%%%%%%%%%%%%%%%%%%%%%%%%%%%%%%%%%%%%%%
%% 2. Observational Data %
%%%%%%%%%%%%%%%%%%%%%%%%%%%%%%%%%%%%%%%%%%%%%%%%%%%%%%%%%
\section{Observational Data}
\label{ObsData}
The AR 12257 appeared on January 6, 2015 near the solar disk center of $E16^oN7^o$ inside an old decaying AR. Its magnetic field emerged continuously at least until January 10, while later on it was too close to the solar limb to monitor the field evolution (it reached the limb on January 14). It had a leading negative polarity as the majority of the ARs located in the northern hemisphere in solar cycle 24. The full evolution of the AR was captured by space-borne Solar Dynamics Observatory as it gives uninterrupted high-cadence full-disk observations of EUV as well as magnetic field. We used the vector magnetic field ($\mathbf{B}$) observations of the \textit{Helioseismic Magnetic Imager}  (HMI; \citealt{schou2012}) at a cadence of 12-minute for five days since the AR emergence. Details of deriving vector magnetic fields from the full-disk filtergrams can be referred to in \citet{bobra2014}, \citet{hoeksema2014}. Figure~\ref{Fig_VMgram} displays representative vector magnetograms of the AR during the emerging phase, which delineates opposite magnetic patches growing both in size and distance. To note, the AR does not have a closed compact polarity inversion line (PIL) as the opposite polarities separate with their emergence. This contrasts with eruptive ARs, e.g., 11158, 11429, 12673 which have intense flares/CMEs \citep{vemareddy2012_hinj,Vemareddy2019_DegEle}. %is not the case with some recent violent 

%    Paragraph: Describe the beginning of B  evolution and relation to magnetic tongue orientation
The evolution of the vertical component of the magnetic field, $B_z$, is characterized by a continuous emergence of bipolar field (Figure~\ref{Fig_VMgram} and associated movie). This emergence started on January 6, 2015 and continues until the AR was too close to the western limb to identify the evolution. The distribution of $B_z$ in both magnetic polarities has extensions so that the magnetogram resembles the yin-yang pattern. These extensions are called magnetic tongues \citep{Lopez2000_CounterRotNonHale, Louni2011_TwistFlxTubeEmer} which present typically in emerging bipolar ARs. They can be understood as the signature of the azimuthal (poloidal) field component of an emerging flux rope. The orientation of the tongues (which is same as the orientation of the PIL) with respect to the line joining the center of the two polarities (dashed line in Figure~\ref{Fig_VMgram}) indicates the sign of magnetic helicity in the emerging flux rope \citep{Poisson2016_PropMagToung}. The evolution of the magnetic tongue pattern in this AR indicates that the emerging field has positive, then negative magnetic helicity (Figure~\ref{Fig_VMgram}).

\begin{figure*}  %---------------------------------------------------------------------
    \centering
	\includegraphics[width=.75\textwidth,clip=]{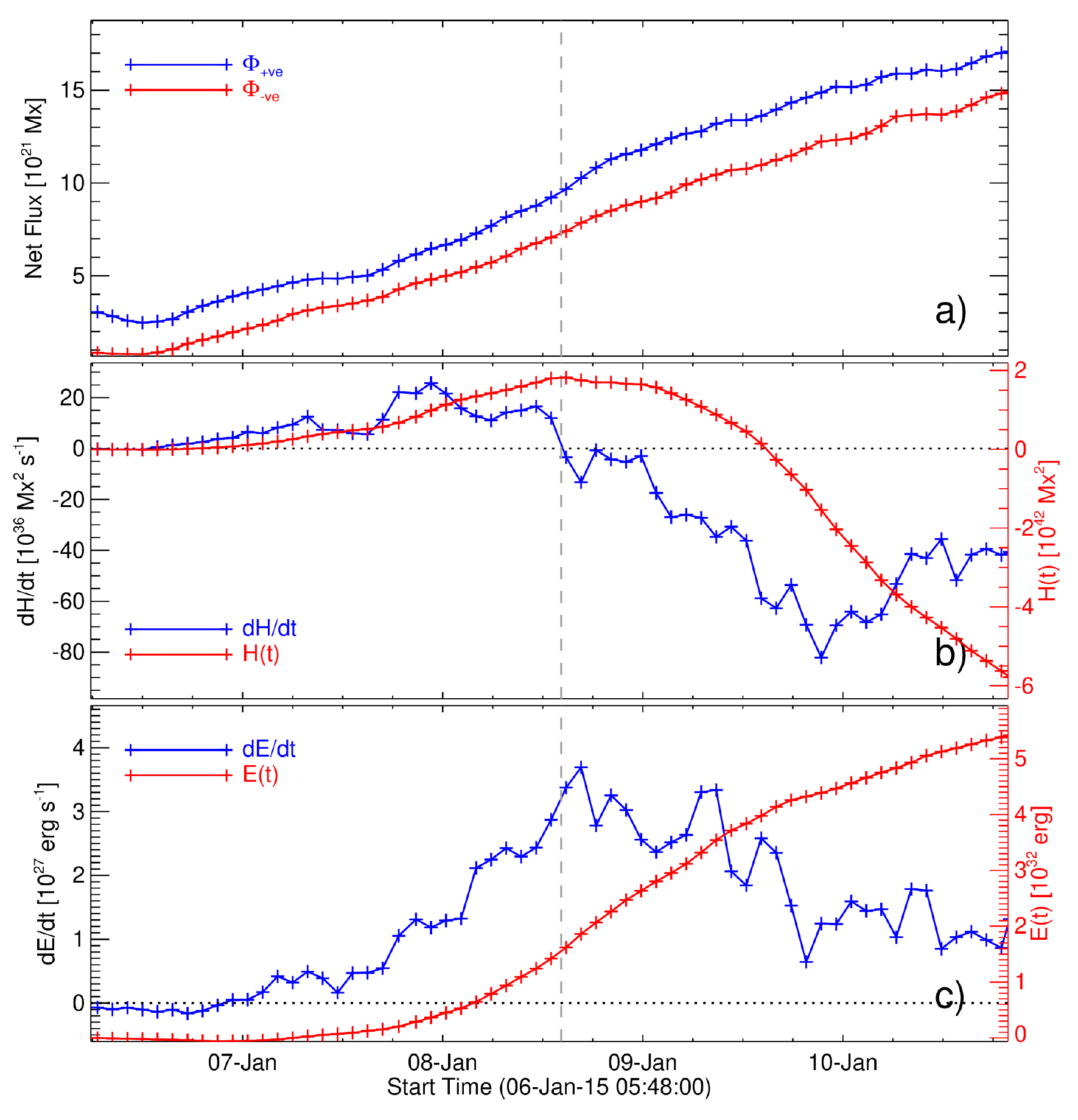}
    \caption{a) Time evolution of the net magnetic flux, b) the helicity injection rate ($dH/dt$) and c) the energy injection rate. Accumulated helicity, $H(t)$, and energy, $E(t)$, are plotted with their scale on the right side axis. Vertical dashed line marks the time (2015-Jan-08T14:00 UT) when $dH/dt$ changes sign from positive to negative.}
    \label{Fig_dhdt}
\end{figure*}

%%%%%%%%%%%%%%%%%%%%%%%%%%%%%%%%%%%%%%%%%%%%%%%%%%%%%%%%%
%% 3. Results %
%%%%%%%%%%%%%%%%%%%%%%%%%%%%%%%%%%%%%%%%%%%%%%%%%%%%%%%%%
\section{Results}
\label{Res}
From the time-sequence vector magnetograms, the velocity field $\mathbf{v}$ is derived from DAVE4VM procedure \citep{schuck2008}. Then the helicity injection rate \citep{Berger1984_Hcons} is computed by
  
  \begin{equation}
  {\left. \frac{dH}{dt} \right|}_{S} =
     2\int_{S}{\left( {{\mathbf{A}}_{p}}\centerdot {{\mathbf{B}}_{t}} \right)  {{\text{v}}_{n}} \,\rmd \mathbf{S}}
    -2 \int_{S}{\left( {{\mathbf{A}}_{p}}\centerdot {{\mathbf{v}}_{t}} \right){{B}_{n}} \,\rmd \mathbf{S}}
  \label{EqHel}
  \end{equation}
where subscript $t$ represents the transverse components and $n$ represents the normal component of $\mathbf{v}$ and $\mathbf{B}$, respectively. $\mathbf{A}_p$ is the vector potential of the coronal potential magnetic field which is derived with $B_n$ as the photospheric boundary condition and with the Coulomb gauge condition, $\nabla \cdot \mathbf{A}_p=0$, in the coronal volume. Details of $dH/dt$ computations can be found in \citet{Liuyang2012_HelEner} and \citet{Vemareddy2015_HelEne}. Similarly magnetic energy injection is computed by 

  \begin{equation}
   {\left. \frac{d{{E}_{m}}}{dt} \right|}_{S} = 
   \frac{1}{4\pi }\int_{S}{B_{t}^{2}{{\text{v}}_{n}}\,\rmd \mathbf{S}} 
  -\frac{1}{4\pi }\int{\left( {{\mathbf{B}}_{t}}\centerdot {{\mathbf{v}}_{t}} \right){{B}_{n}}\,\rmd \mathbf{S}}
  \label{EqEne}
  \end{equation}

In Figure~\ref{Fig_dhdt}, we plot the time evolution of net magnetic flux $\Phi =\sum\limits_{pix}^{{}}{{{B}_{z}}\,\Delta x\Delta y}$, the helicity injection rate $dH/dt$ and the energy injection rate $dE/dt$. $dH/dt$ exhibits a change of sign from positive to negative around mid of January 8. $dH/dt$ of positive (negative) sign indicates positive (negative) chirality of the magnetic field in the AR magnetic system. 
This is related to the change in the magnetic tongue pattern described in the previous section. More quantitatively, the flux of helicity in Equation~(\ref{EqHel}) can be rewritten with the horizontal (tangential) photospheric velocity $\mathbf{u}$ of the elementary magnetic flux tube defined as \citep{Demoulin2003_MagEneHel}

   \begin{equation}
    \mathbf{u} = \mathbf{v}_t-\frac{v_n}{B_n} \mathbf{B}_t
    \label{EqCC}
   \end{equation}
   
Then Equation~(\ref{EqHel}) is simplified by replacing $\mathbf{V}_t$ by $\mathbf{u}$ and with $v_n=0$. This formulation shows that the evolution of the $B_n$ magnetogram allows to derive the helicity flux from the magnetic polarity motion ($\mathbf{u}$). This links the evolution of the magnetic tongues to the magnetic helicity injection. Next, since there is more negative helicity injected later on, the accumulated helicity in the corona ( $H(t)=\int_{t'=0}^{t} \frac{dH}{dt} \rmd t'$ ) also turns sign from positive to negative by the end of January 9. 

\begin{figure*} %---------------------------------------------------------------------
    \centering
	\includegraphics[width=.75\textwidth,clip=]{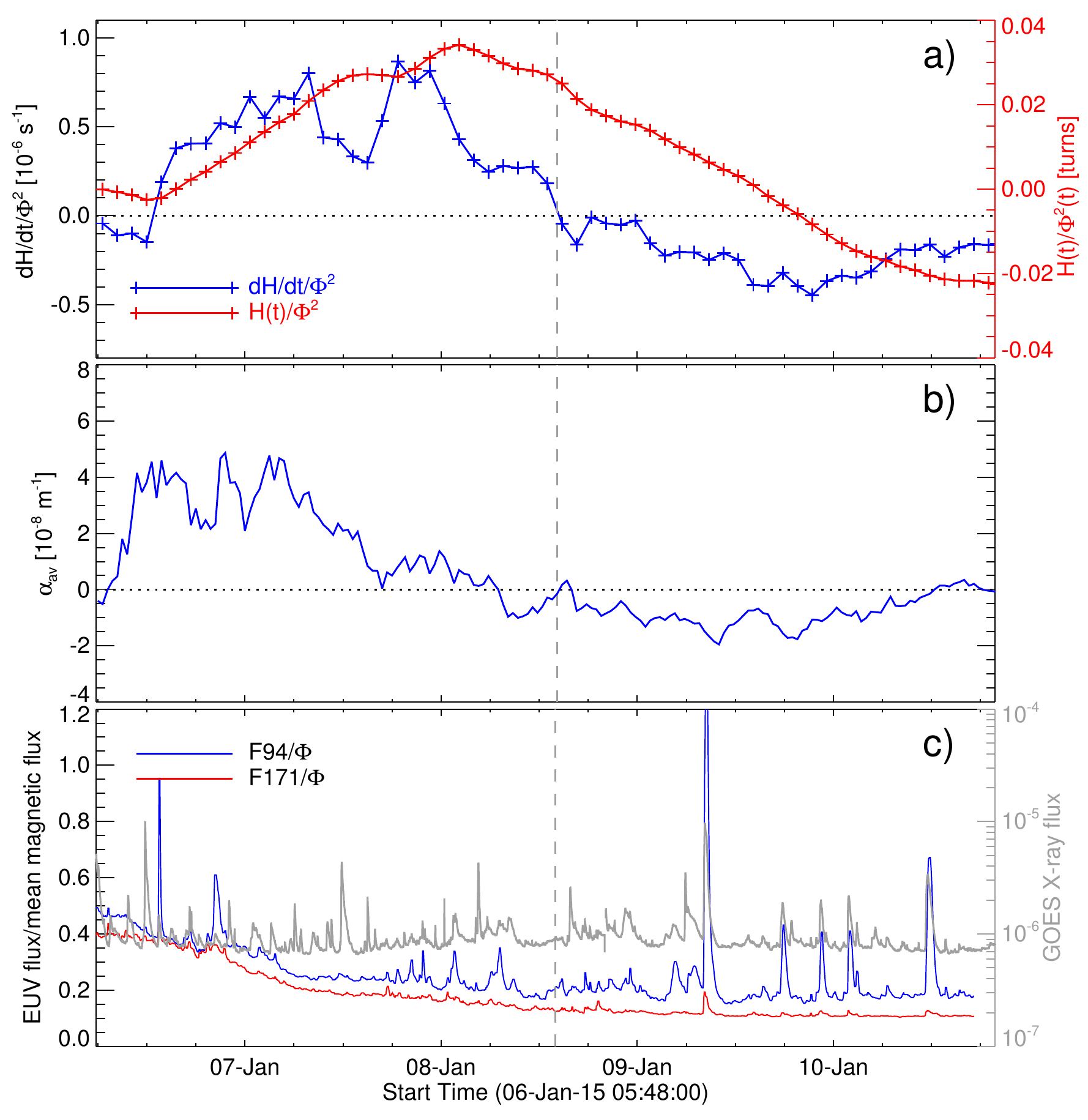}
    \caption{a) Time-evolution of $dH/dt$ normalized by the square of the mean magnetic flux, $\Phi (t)$. $H(t)/\Phi^2 (t)$ is also shown with y-axis scale on right. Its variation is within 0.03 turn with a sign change marked by the vertical dashed line.
     b) Evolution of the mean twist parameter $\alpha_{av}$, whose sign change is also consistent with $dH/dt/\Phi^2$. 
     c) Light curves of the AR in EUV intensity 94 and 171 \AA\ wavebands divided by $\Phi (t)$. GOES flux is also shown with y-axis scale on right. Prominent peaks in EUV 94~\AA~during January 9-10 are co-temporal with X-ray peaks of C-class magnitude flares.   }
    \label{Fig_dhdt_phi2}
\end{figure*}

In contrast, the energy flux $dE/dt$ (Figure~\ref{Fig_dhdt}c), is always positive (except marginally at the beginning). It first grows before declining after 2 days of emergence. $dE/dt$ is of the order of $10^{27}$ erg/s. This implies that the coronal magnetic energy content becomes $10^{32}$ erg by January 8.  Only a fraction of this energy, the one in excess of the potential field energy computed from the same $B_n$ magnetogram, is available to power the activity. From the GOES and EUV data analysis, the AR produced only C-class flares (Figure~\ref{Fig_dhdt_phi2}c).

\begin{figure*} %---------------------------------------------------------------------
    \centering
	\includegraphics[width=.9\textwidth,clip=]{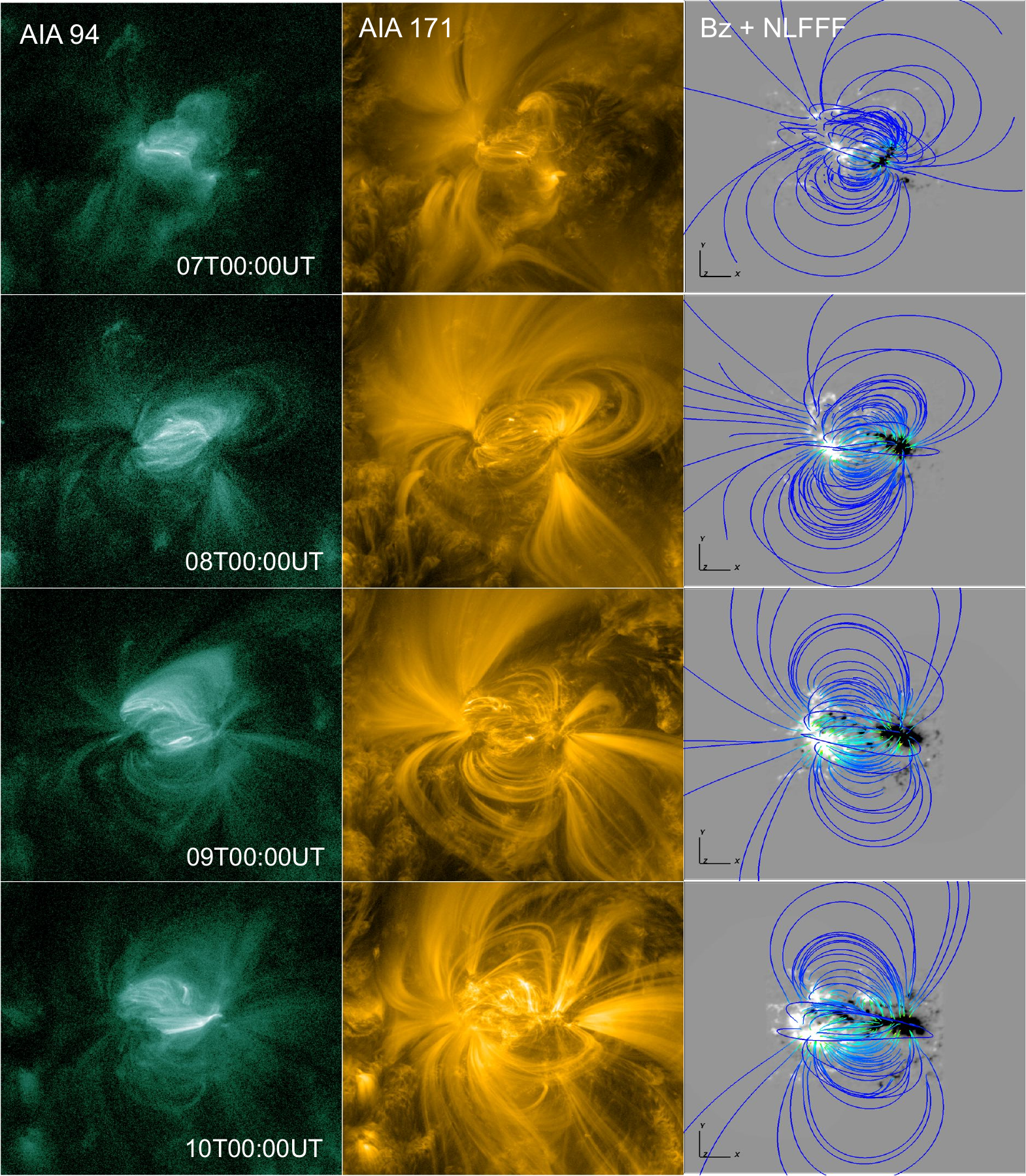}
    \caption{Comparison of NLFFF magnetic structure with plasma tracers in EUV images. AIA 94 \AA~(first column), 171 \AA~(2nd column) images at different epochs of the AR evolution. The third column is the NLFFF magnetic structure obtained at the same time of EUV observations. The background image is the vertical component, $B_n$, map overlaid by traced field lines. Globally, the modelled magnetic structure resembles the EUV plasma loops to a large extent. The field of view is the same in all panels. To support this figure, a movie is available in the Electronic Supplementary Materials. The movie shows the full evolution of the AR in the simultaneous images of AIA 94\AA, 304\AA, 171\AA, and HMI line-of-sight magnetic field component. Start (end) time of the movie is 2015-01-06T05:30 UT (2015-01-10T15:00 UT).
    }
    \label{Fig_extp}
\end{figure*}  

In Figure~\ref{Fig_dhdt_phi2}a, we plot the helicity flux normalized by the mean magnetic flux of the magnetic polarities ($ \Phi =( \Phi_+ +|\Phi_-|)/2)$.  This normalized quantity, $\frac{dH}{dt}/\Phi ^{2} (t)$, measures the helicity flux per unit flux. This allows to compare the helicity flux among ARs having different $\Phi$ magnitude.  $|\frac{dH}{dt}/{{\Phi }^{2}}|$ stays below $ 0.8\times 10^{-6} $ s$^{-1}$ which is modest for an AR. We also normalize the accumulated helicity by $\Phi^2$ value.  $H(t)/\Phi^2 (t)$ reaches 0.03 turn and then changes sign to -0.02 turn. This is a modest input by a comparison with AR 12371, which produces 4 eruptions, since this value was found to be 0.15 turn \citep{Vemareddy2017_homologus}. 

We next compute the mean force-free twist parameter 
$\alpha_{av} = \int J_n \, {\rm sign}(B_n) \,\rmd \mathbf{S} / \int |B_n| \,\rmd \mathbf{S}$. $\alpha_{av}$ measures the non-potentiality of the AR magnetic field, with $\alpha$ being constant along any field line of a force-free field. $\alpha_{av}$ also evolves from positive to negative values with a change of sign at the beginning of January 8 (Figure~\ref{Fig_dhdt_phi2}b). This is another indication that the coronal field changed helicity sign, as the signs of magnetic and current helicity, then of $\alpha_{av}$, are typically expected in coronal configurations \citep{Russell2019}.
% Further, the change of sign in $dH/dt$ is verified with the , which is plotted in Figure~\ref{Fig_dhdt_phi2}b. The $\alpha_{av}$ measures average twist or shearedness of the AR magnetic field, whose change of sign from +ve to -ve is co-temporal with that of $dH/dt$.    

Finally, we derived the light curves of the EUV emission of the AR in AIA 94 and 171~\AA. For this purpose, we use AIA observations of the AR at 3 minute cadence. % for the entire time period of study. 
The EUV flux in ARs is typically proportional to the magnetic flux \citep[e.g.][]{Demoulin2004}. Since the AR has a increasing magnetic flux, we normalized the light curves with the mean magnetic flux of the opposite polarities and plotted them in Figure~\ref{Fig_dhdt_phi2}c. Two flares are registered in AIA 94 on January 6 which are not identified by the GOES X-ray flux. Next, during the period of $dH/dt<0$, on January 9 and 10, five C-class flares occurred (the largest is a C9.6 on 09/08:04 UT). By a careful inspection, we noted that none of these flares are associated with CMEs visible in LASCO C2 field of view.  

The change of sign of the photospheric magnetic helicity flux with time has implications on the global magnetic configuration of the AR. %In order to provide clues about the coronal field evolution we next compute the coronal field, so that to locate where opposite helicities are located. 
In order to realize the change in the magnetic configuration, we modelled the coronal magnetic field of the AR by non-linear force-free field (NLFFF) extrapolation technique \citep{Wheatland2000,wiegelmann2004,wiegelmann2010}. The photospheric vector magnetic observations $\mathbf{B}$ are employed as lower boundary conditions and are embedded in an extended region by padding with null array. The final computations are carried out in a Cartesian grid of $400\times400\times256$ corresponding to a physical volume of $292\times292\times 187$ Mm$^3$. 

In Figure~\ref{Fig_extp}, the rendered magnetic structure at different times is compared with the coronal plasma loops present in the images of AIA 94 and 171 \AA. Plasma emission in 94 \AA\ waveband captures compact loops of the AR, whereas the 171~\AA\,waveband shows higher lying loops together gives a more global picture of the AR magnetic structure. The modelled magnetic structure, to a great extent, resembles the morphology of plasma loops in AIA passbands. On January 6-7, the core of the AR is formed by an arcade of coronal loops with a positive magnetic shear (as deduced from their crossing angle with the PIL). This is coherent with the field line orientations and with $\alpha_{av}$  (Figure~\ref{Fig_dhdt_phi2}b). In between the end of January 7 and the beginning of January 8, the observed coronal loops have no marked shear, and the extrapolated field is nearly potential (second row of Figure~\ref{Fig_extp}), in agreement with the low value of $\alpha_{av}$. By mid January 8, the coronal arcade in the core of the AR has a marked negative shear, and this is observed later on. This is again in agreement of the computed field lines, $dH/dt$, and $\alpha_{av}$.

Having 3D magnetic field of the AR corona, we can also estimate the relative helicity with respect a reference field $\mathbf{B}_p$ which is having the same normal field component as that of the magnetic field being evaluated \citep{Berger1984} 
\begin{equation}
H_R=\int_V (\mathbf{A}-\mathbf{A_p})\cdot(\mathbf{B}-\mathbf{B}_p) dV 
\end{equation}
where $\mathbf{A}$ and $\mathbf{A}_p$ are vector potentials of $\mathbf{B}$ and $\mathbf{B}_p$ respectively, which are constructed with Devore guage condition \citep{DeVore2000_HdiffRot,Valori2016}. The $H_R$ is estimated to be [0.40, 1.01, -2.43, -3.17]$\times10^{42}$ Mx$^2$ respectively for the observations on Jan 7, 8, 9 and 10. To some approximation, the first two values are consistent with the values derived from time integration of $dH/dt$, but the third value is way different in sign itself. Two approaches differ in the values because the later lacks information on the coronal helicity that was injected through the photospheric surface and particularly in this case the sign change in $dH/dt$ adds to the problem of estimating coronal helicity from time-integration approach.

%Before the time of the sign reversal of $dH/dt$ (about January 8, 12:00 UT), the global structure appears as a forward sigmoid which is consistent with the positive sign of the injected helicity. In the later time, the magnetic structure evolves to become inverse sigmoid (inverse-S) as the sign of $dH/dt$ changes to negative. 

The above coronal evolution is also consistent with the injection of magnetic helicity shown in Figure~\ref{Fig_dhdt}b as follows. There is a clear decrease of the mean helicity flux at the beginning of January 8 which is due to the emergence of some negative helicity, while the total flux is still positive. This is associated to the retraction of the tongues indicating positive helicity and the tongues increasingly become reverse orientation. Since the magnetic flux, which succeeds to cross the photosphere, typically expands in the corona in numerical simulations, in agreement with observations, the corona has to respond to the changing boundary condition under the scenario of quasi-static evolution \citep{vemareddy2014_Quasi_Stat}. As a result, the coronal magnetic structure at each time is close to an equilibrium corresponding to the evolution of the observed boundary field. 

The above photospheric and coronal observations, in agreement with the NLFFF extrapolation, provide a clear description of this unusual AR evolution where opposite magnetic helicities are successively injected by a continuous emergence of a bipolar magnetic field. Since the presence of finite magnetic helicity limits the available free energy well above the potential field energy, the presence of opposite magnetic helicities is a favorable condition for the coronal field to be able to release more magnetic energy, for a given initial and potential energies. Indeed, a possible enhanced flaring activity was suggested to be a result of cancellation of helicity of the oppositely sheared field as shown by numerical simulations of \citep{Kusano2004_RevShArc}. 

 % Little discussion for CME-less case;
Moreover, this AR also does not develop a sigmoidal shape in EUV (Figure \ref{Fig_extp}), as observed in many ARs as the consequence of reconnection at the PIL of sheared loops \citep{green2009, green2011, savcheva2012a}. Then, in this AR the diffusion of the magnetic polarities, then cancellation at the PIL, followed by the build up of a flux rope, as found in MHD simulations \citep{amari2003a, aulanier2010}, is not efficient enough. Indeed, the studied AR has intense magnetic field of opposite magnetic polarities well separated on both sides of the PIL and continuous emergence, with separation of the polarities, prevent flux cancellation to occur at the PIL in this emerging phase. In the context of previous observations and numerical simulations, this observed evolution do not favor the formation of a coronal flux rope which is indeed not found with the extrapolated magnetic field. 

%%%%%%%%%%%%%%%%%%%%%%%%%%%%%%%%%%%%%%%%%%%%%%%%%%%%%%%%%
%% 4. Summary and Discussion %
%%%%%%%%%%%%%%%%%%%%%%%%%%%%%%%%%%%%%%%%%%%%%%%%%%%%%%%%%
\section{Summary and Discussion}
\label{SumDisc}

% Summary of CME-less AR with sign-change dH/dt --touched with interpretation 
We studied an uncommon AR with photospheric motions generating successive injection of opposite sign of magnetic helicity within a continuous process of magnetic emergence. The origins of this sign change lies in the sub-photospheric formation of this coherent magnetic structure having opposite magnetic helicities. These are challenging observations for dynamo models. A proxy of the helicity injection is the most easily seen in longitudinal magnetograms with a yin-yang pattern of the bipolar field, or polarity elongations called magnetic tongues. While in most emerging bipolar ARs, the magnetic tongues indicate injection of magnetic helicity of a given sign, in the studied AR, the yin-yang pattern reverse its orientation during the emergence, indicating a change of helicity sign injection. This reversal is confirmed by the computation of the magnetic helicity flux. This is also coherent with the evolution of the mean force-free parameter, $\alpha$, and of the observed shear of coronal loops.

%distribution of the magnetic tongue evolution, so the evolution of the photospheric magnetic field distribution. A dominant sign of positive (negative) $dH/dt$ is due to a large fraction of the magnetogram injecting positive helicity flux. 

%  Reasons for flaring, heating in successive opposite dH/dt ARs
Earlier studies reported a tendency that mixed signs of helicity flux is found in flare-producing ARs, whereas CME producing regions have injection of a predominant sign (either positive or negative) helicity flux \citep{MoonYJ2002_ImpulsiveVar,Romano2011_SolEru_Trig_OppHel,vemareddy2012_hinj,vemareddy2012_sunspot_rot}. Based on these kinds of observations, \citet{Kusano2003_AnnihMagHel} suggested that coexistence of positive and negative helicities are important for the onset of flares and further proposed a numerical model to convert a sheared field to shear-free by magnetic reconnection of oppositely sheared magnetic fields. In the studied AR, a similar evolution is present with $dH/dt$ changing sign from positive to negative. The origins of this sign change lies in the distribution of $dH/dt$. A dominant sign of positive (negative) $dH/dt$ is due to a large fraction of the magnetogram injecting positive helicity flux for a given foot point motions. When the initial positive H-flux is replaced gradually by negative H-flux in the later part of the evolution, the loop system in the AR also transforms shear from positive to negative sign. In this process, current sheet forms between opposite sheared field. Then reconnection of those field lines gives rise to intermittent flares of any magnitude depending on the local shear distribution, as observed in this AR. It is also very likely that the reconnection between the opposite helicities occurs continuously in the interacting magnetic fields with different connectivities, then the slow heating causes enhanced EUV emission. Therefore, the reconnection between opposite sheared field is intermittent to a large scale and gradual to a small scale field. Based on the observations in AR 11928, \citet{Vemareddy2017_SucHelInj} suggested that the cancellation of the helicity during the evolution leads to relaxation of the sheared field by magnetic reconnection causing small-scale flares and coronal heating. 

% About coronal Helicity Accumulation and CMEs 
From the helicity accumulation scenario, ARs with change of $dH/dt$ sign may not favor the twisted flux rope formation and have smaller content of coronal helicity below the eruptive threshold, so a CME eruption is unlikely \citep{zhangmei2005, Zhangmei2006_MagFld_confin}. Investigations by \citet{Zhangmei2008_HelBound} revealed that the magnitude of the helicity upper bound of force-free fields is non-trivially dependent on the boundary condition. Their study indicated that the dipolar field had a helicity (helicity value normalized by the square of the respective surface poloidal flux equivalent to $H(t)/\Phi^2(t)$) upper bound of 0.35 which is comparable to the observational estimations of \citet{Demoulin2007_adsr} and the recent emerging active regions with CME occurrences reported in \citet{Vemareddy2015_HelEne}. Importantly, for the multipolar boundary condition \citet{Zhangmei2008_HelBound} found a helicity upper bound 10 times smaller than those with a dipolar boundary condition. The present studied AR 12257 is having an overall bipolar configuration with $H(t)/\Phi^2$ well below 0.035 turns (See Figure~\ref{Fig_dhdt_phi2}a). Another such example AR 11928 of opposite sign $dH/dt$ also have this value 0.07 which is far below the theoretical upper bound \citep{Vemareddy2017_SucHelInj}. Therefore, the ARs with reversal of the helicity flux are arguably with coronal helicity below the upper bound and are very important to identify CME-less (but confined flare productive) ARs readily. 
	
Further, we would point that the theoretical estimations of helicity upper bound are still debated because several eruptive ARs were not differentiable from their counterparts with that value \citep{Labonte2007_SurveyMagHel, Demoulin2007_helicity_adsr, Demoulin2009_adsr}. It remains to be explored further for the precise value of the helicity upper bound, by assessing the ARs of eruptive and non-eruptive behavior based on the statistical value together with the twisted flux in the magnetic structure. In the case of not having an idea of this limit, we can directly probe the twisted flux, as the key information, in the AR magnetic structure. 

% Final verdict for ARs to be CME productive
Even continuous injection of $dH/dt$ may not ensure occurrence of a CME.  While accumulation of magnetic helicity of a given sign leads to the helicity upper bound but it is not a sufficient condition. Such ARs have also been observed recently. The AR 12192 is one such example which is flare productive with large X-class flares none of them are associated with CMEs \citep{SunX2015_WhyFlareRichCMELess, Vemareddy2017_homologus}. The $dH/dt$ evolution in 12192 is a positive sign during its entire evolution. However, the extrapolated field does not show the presence of a twisted flux rope. So the eruptive nature of an AR is decided by two factors, 1) continuous injection of a given sign of helicity by the AR magnetic field, 2) formation and existence of twisted flux rope.  \citet{vemareddy2019_VeryFast} showed that $dH/dt/\Phi^2$ for CME-less AR has a smaller (factor ten) helicity injection rate than the CME producing ARs, which needs to be verified in larger sample of ARs. From the point of space-weather, the characteristics of helicity flux evolution in ARs is very much important to assess the problem of coronal helicity threshold for the eruptivity of the magnetic field. 

% Time of dH/dt reversal : 2015-01-08T14:00

\section*{Acknowledgements} SDO is a mission of NASA's Living With a Star Program. The author is grateful to P.~D\'emoulin for a detailed suggestions especially the discussion on magnetic tongue concept. I thank an anonymous referee for encouraging comments and suggestions.

\section*{Data Availability} The data used in this manuscript is from NASA's SDO mission and is publicly available from Joint Science Operations Center (http://jsoc.stanford.edu/).

\bibliographystyle{mnras.bst}
%\bibliography{../ref_lib18032021.bib}

\begin{thebibliography}{}
	\makeatletter
	\relax
	\def\mn@urlcharsother{\let\do\@makeother \do\$\do\&\do\#\do\^\do\_\do\%\do\~}
	\def\mn@doi{\begingroup\mn@urlcharsother \@ifnextchar [ {\mn@doi@}
		{\mn@doi@[]}}
	\def\mn@doi@[#1]#2{\def\@tempa{#1}\ifx\@tempa\@empty \href
		{http://dx.doi.org/#2} {doi:#2}\else \href {http://dx.doi.org/#2} {#1}\fi
		\endgroup}
	\def\mn@eprint#1#2{\mn@eprint@#1:#2::\@nil}
	\def\mn@eprint@arXiv#1{\href {http://arxiv.org/abs/#1} {{\tt arXiv:#1}}}
	\def\mn@eprint@dblp#1{\href {http://dblp.uni-trier.de/rec/bibtex/#1.xml}
		{dblp:#1}}
	\def\mn@eprint@#1:#2:#3:#4\@nil{\def\@tempa {#1}\def\@tempb {#2}\def\@tempc
		{#3}\ifx \@tempc \@empty \let \@tempc \@tempb \let \@tempb \@tempa \fi \ifx
		\@tempb \@empty \def\@tempb {arXiv}\fi \@ifundefined
		{mn@eprint@\@tempb}{\@tempb:\@tempc}{\expandafter \expandafter \csname
			mn@eprint@\@tempb\endcsname \expandafter{\@tempc}}}
	
	\bibitem[\protect\citeauthoryear{{Amari}, {Luciani}, {Aly}  \& {et al}}{{Amari}
		et~al.}{2003}]{amari2003a}
	{Amari} T.,  {Luciani} J.~F.,  {Aly} J.~J.,   {et al} 2003, \mn@doi [\apj]
	{10.1086/345501}, \href {http://adsabs.harvard.edu/abs/2003ApJ...585.1073A}
	{585, 1073}
	
	\bibitem[\protect\citeauthoryear{{Aulanier}, {T{\"o}r{\"o}k}, {D{\'e}moulin}
		\& {DeLuca}}{{Aulanier} et~al.}{2010}]{aulanier2010}
	{Aulanier} G.,  {T{\"o}r{\"o}k} T.,  {D{\'e}moulin} P.,   {DeLuca} E.~E.,
	2010, \mn@doi [\apj] {10.1088/0004-637X/708/1/314}, \href
	{http://adsabs.harvard.edu/abs/2010ApJ...708..314A} {708, 314}
	
	\bibitem[\protect\citeauthoryear{{Berger}}{{Berger}}{1984}]{Berger1984_Hcons}
	{Berger} M.~A.,  1984, \mn@doi [Geophysical and Astrophysical Fluid Dynamics]
	{10.1080/03091928408210078}, \href
	{http://adsabs.harvard.edu/abs/1984GApFD..30...79B} {30, 79}
	
	\bibitem[\protect\citeauthoryear{Berger \& Field}{Berger \&
		Field}{1984}]{Berger1984}
	Berger M.~A.,  Field G.~B.,  1984, \mn@doi [Journal of Fluid Mechanics]
	{10.1017/S0022112084002019}, 147, 133
	
	\bibitem[\protect\citeauthoryear{{Bobra}, {Sun}, {Hoeksema}  \& {et
			al}}{{Bobra} et~al.}{2014}]{bobra2014}
	{Bobra} M.~G.,  {Sun} X.,  {Hoeksema} J.~T.,   {et al} 2014, \mn@doi [\solphys]
	{10.1007/s11207-014-0529-3}, \href
	{http://adsabs.harvard.edu/abs/2014SoPh..289.3549B} {289, 3549}
	
	\bibitem[\protect\citeauthoryear{{Chae}}{{Chae}}{2001}]{Chae2001_ObsDetMagHel}
	{Chae} J.,  2001, \mn@doi [\apjl] {10.1086/324173}, \href
	{https://ui.adsabs.harvard.edu/abs/2001ApJ...560L..95C} {560, L95}
	
	\bibitem[\protect\citeauthoryear{{DeVore}}{{DeVore}}{2000}]{DeVore2000_HdiffRot}
	{DeVore} C.~R.,  2000, \mn@doi [\apj] {10.1086/309274}, \href
	{https://ui.adsabs.harvard.edu/abs/2000ApJ...539..944D} {539, 944}
	
	\bibitem[\protect\citeauthoryear{{D{\'e}moulin}}{{D{\'e}moulin}}{2004}]{Demoulin2004}
	{D{\'e}moulin} P.,  2004, in {Stepanov} A.~V.,  {Benevolenskaya} E.~E.,
	{Kosovichev} A.~G.,  eds,  Vol. 223, Multi-Wavelength Investigations of Solar
	Activity. pp 13--22, \mn@doi{10.1017/S1743921304005046}
	
	\bibitem[\protect\citeauthoryear{{D{\'e}moulin}}{{D{\'e}moulin}}{2007}]{Demoulin2007_helicity_adsr}
	{D{\'e}moulin} P.,  2007, \mn@doi [Advances in Space Research]
	{10.1016/j.asr.2006.12.037}, \href
	{https://ui.adsabs.harvard.edu/abs/2007AdSpR..39.1674D} {39, 1674}
	
	\bibitem[\protect\citeauthoryear{{D{\'e}moulin} \& {Berger}}{{D{\'e}moulin} \&
		{Berger}}{2003}]{Demoulin2003_MagEneHel}
	{D{\'e}moulin} P.,  {Berger} M.~A.,  2003, \solphys, \href
	{http://adsabs.harvard.edu/abs/2003SoPh..215..203D} {215, 203}
	
	\bibitem[\protect\citeauthoryear{D{\'e}moulin \& Pariat}{D{\'e}moulin \&
		Pariat}{2009}]{Demoulin2009_adsr}
	D{\'e}moulin P.,  Pariat E.,  2009, \mn@doi [Advances in Space Research]
	{https://doi.org/10.1016/j.asr.2008.12.004}, 43, 1013
	
	\bibitem[\protect\citeauthoryear{{D{\'e}moulin}, {Mandrini}, {Van
			Driel-Gesztelyi}  \& {et al}}{{D{\'e}moulin}
		et~al.}{2002a}]{Demoulin2002_MagHelShMot}
	{D{\'e}moulin} P.,  {Mandrini} C.~H.,  {Van Driel-Gesztelyi} L.,   {et al}
	2002a, \mn@doi [\solphys] {10.1023/A:1015531804337}, \href
	{http://adsabs.harvard.edu/abs/2002SoPh..207...87D} {207, 87}
	
	\bibitem[\protect\citeauthoryear{{D{\'e}moulin}, {Mandrini}, {van
			Driel-Gesztelyi}  \& {et al}}{{D{\'e}moulin}
		et~al.}{2002b}]{Demoulin2002_Hbudget}
	{D{\'e}moulin} P.,  {Mandrini} C.~H.,  {van Driel-Gesztelyi} L.,   {et al}
	2002b, \mn@doi [\aap] {10.1051/0004-6361:20011634}, \href
	{https://ui.adsabs.harvard.edu/abs/2002A&A...382..650D} {382, 650}
	
	\bibitem[\protect\citeauthoryear{Dhakal, Zhang, Vemareddy  \& Karna}{Dhakal
		et~al.}{2020}]{DhakalS2020}
	Dhakal S.~K.,  Zhang J.,  Vemareddy P.,   Karna N.,  2020, \mn@doi [\apj]
	{10.3847/1538-4357/abacbc}, 901, 40
	
	\bibitem[\protect\citeauthoryear{{Green} \& {Kliem}}{{Green} \&
		{Kliem}}{2009}]{green2009}
	{Green} L.~M.,  {Kliem} B.,  2009, \mn@doi [\apjl]
	{10.1088/0004-637X/700/2/L83}, \href
	{http://adsabs.harvard.edu/abs/2009ApJ...700L..83G} {700, L83}
	
	\bibitem[\protect\citeauthoryear{{Green}, {Kliem}  \& {Wallace}}{{Green}
		et~al.}{2011}]{green2011}
	{Green} L.~M.,  {Kliem} B.,   {Wallace} A.~J.,  2011, \mn@doi [\aap]
	{10.1051/0004-6361/201015146}, \href
	{http://adsabs.harvard.edu/abs/2011A%26A...526A...2G} {526, A2}
		
	\bibitem[\protect\citeauthoryear{{Hoeksema}, {Liu}, {Hayashi}, {Sun}  \& {et
				al}}{{Hoeksema} et~al.}{2014}]{hoeksema2014}
		{Hoeksema} J.~T.,  {Liu} Y.,  {Hayashi} K.,  {Sun} X.,   {et al} 2014, \mn@doi
		[\solphys] {10.1007/s11207-014-0516-8}, \href
		{http://adsabs.harvard.edu/abs/2014SoPh..289.3483H} {289, 3483}
		
	\bibitem[\protect\citeauthoryear{{Kusano}, {Maeshiro}, {Yokoyama}  \&
			{Sakurai}}{{Kusano} et~al.}{2002}]{Kusano2002}
		{Kusano} K.,  {Maeshiro} T.,  {Yokoyama} T.,   {Sakurai} T.,  2002, \mn@doi
		[\apj] {10.1086/342171}, \href
		{https://ui.adsabs.harvard.edu/abs/2002ApJ...577..501K} {577, 501}
		
	\bibitem[\protect\citeauthoryear{{Kusano}, {Yokoyama}, {Maeshiro}  \&
			{Sakurai}}{{Kusano} et~al.}{2003}]{Kusano2003_AnnihMagHel}
		{Kusano} K.,  {Yokoyama} T.,  {Maeshiro} T.,   {Sakurai} T.,  2003, \mn@doi
		[Advances in Space Research] {10.1016/S0273-1177(03)90628-4}, \href
		{https://ui.adsabs.harvard.edu/abs/2003AdSpR..32.1931K} {32, 1931}
		
	\bibitem[\protect\citeauthoryear{{Kusano}, {Maeshiro}, {Yokoyama}  \&
			{Sakurai}}{{Kusano} et~al.}{2004}]{Kusano2004_RevShArc}
		{Kusano} K.,  {Maeshiro} T.,  {Yokoyama} T.,   {Sakurai} T.,  2004, \mn@doi
		[\apj] {10.1086/421547}, \href
		{https://ui.adsabs.harvard.edu/abs/2004ApJ...610..537K} {610, 537}
		
	\bibitem[\protect\citeauthoryear{{LaBonte}, {Georgoulis}  \& {Rust}}{{LaBonte}
			et~al.}{2007}]{Labonte2007_SurveyMagHel}
		{LaBonte} B.~J.,  {Georgoulis} M.~K.,   {Rust} D.~M.,  2007, \mn@doi [\apj]
		{10.1086/522682}, \href
		{https://ui.adsabs.harvard.edu/abs/2007ApJ...671..955L} {671, 955}
		
	\bibitem[\protect\citeauthoryear{{Liu} \& {Schuck}}{{Liu} \&
			{Schuck}}{2012}]{Liuyang2012_HelEner}
		{Liu} Y.,  {Schuck} P.~W.,  2012, \mn@doi [\apj] {10.1088/0004-637X/761/2/105},
		\href {http://adsabs.harvard.edu/abs/2012ApJ...761..105L} {761, 105}
		
	\bibitem[\protect\citeauthoryear{Liu, Hoeksema, Bobra, Hayashi, Schuck  \&
			Sun}{Liu et~al.}{2014}]{LiuYang2014_MagHelEmerARs}
		Liu Y.,  Hoeksema J.~T.,  Bobra M.,  Hayashi K.,  Schuck P.~W.,   Sun X.,
		2014, \mn@doi [\apj] {10.1088/0004-637x/785/1/13}, 785, 13
		
	\bibitem[\protect\citeauthoryear{{L{\'o}pez Fuentes}, {Demoulin}, {Mandrini}
			\& {van Driel-Gesztelyi}}{{L{\'o}pez Fuentes}
			et~al.}{2000}]{Lopez2000_CounterRotNonHale}
		{L{\'o}pez Fuentes} M.~C.,  {Demoulin} P.,  {Mandrini} C.~H.,   {van
			Driel-Gesztelyi} L.,  2000, \mn@doi [\apj] {10.1086/317180}, \href
		{https://ui.adsabs.harvard.edu/abs/2000ApJ...544..540L} {544, 540}
		
	\bibitem[\protect\citeauthoryear{{Luoni}, {D{\'e}moulin}, {Mandrini}  \& {van
				Driel-Gesztelyi}}{{Luoni} et~al.}{2011}]{Louni2011_TwistFlxTubeEmer}
		{Luoni} M.~L.,  {D{\'e}moulin} P.,  {Mandrini} C.~H.,   {van Driel-Gesztelyi}
		L.,  2011, \mn@doi [\solphys] {10.1007/s11207-011-9731-8}, \href
		{https://ui.adsabs.harvard.edu/abs/2011SoPh..270...45L} {270, 45}
		
	\bibitem[\protect\citeauthoryear{{Moffatt}}{{Moffatt}}{1978}]{Mofatt1978_book}
		{Moffatt} H.~K.,  1978, {Magnetic field generation in electrically conducting
			fluids}.
		Cambridge: University Press
		
	\bibitem[\protect\citeauthoryear{{Moon}, {Chae}, {Wang}, {Choe}  \&
			{Park}}{{Moon} et~al.}{2002}]{MoonYJ2002_ImpulsiveVar}
		{Moon} Y.~J.,  {Chae} J.,  {Wang} H.,  {Choe} G.~S.,   {Park} Y.~D.,  2002,
		\mn@doi [\apj] {10.1086/343130}, \href
		{https://ui.adsabs.harvard.edu/abs/2002ApJ...580..528M} {580, 528}
		
	\bibitem[\protect\citeauthoryear{{Moraitis}, {Sun}, {Pariat}  \&
			{Linan}}{{Moraitis} et~al.}{2019}]{Moraitis2019_MagHel_erup}
		{Moraitis} K.,  {Sun} X.,  {Pariat} {\'E}.,   {Linan} L.,  2019, \mn@doi [\aap]
		{10.1051/0004-6361/201935870}, \href
		{https://ui.adsabs.harvard.edu/abs/2019A&A...628A..50M} {628, A50}
		
	\bibitem[\protect\citeauthoryear{{Pariat}, {D{\'e}moulin}  \&
			{Berger}}{{Pariat} et~al.}{2005}]{pariat2005}
		{Pariat} E.,  {D{\'e}moulin} P.,   {Berger} M.~A.,  2005, \mn@doi [\aap]
		{10.1051/0004-6361:20052663}, \href
		{http://adsabs.harvard.edu/abs/2005A%26A...439.1191P} {439, 1191}
			
	\bibitem[\protect\citeauthoryear{{Pariat}, {Valori}, {D{\'e}moulin}  \&
				{Dalmasse}}{{Pariat} et~al.}{2015}]{Pariat2015_TestMagHelConsr}
			{Pariat} E.,  {Valori} G.,  {D{\'e}moulin} P.,   {Dalmasse} K.,  2015, \mn@doi
			[\aap] {10.1051/0004-6361/201525811}, \href
			{https://ui.adsabs.harvard.edu/abs/2015A&A...580A.128P} {580, A128}
			
	\bibitem[\protect\citeauthoryear{{Park}, {Chae}  \& {Wang}}{{Park}
				et~al.}{2010}]{Park2010_ProdFlareMagHel}
			{Park} S.-h.,  {Chae} J.,   {Wang} H.,  2010, \mn@doi [\apj]
			{10.1088/0004-637X/718/1/43}, \href
			{https://ui.adsabs.harvard.edu/abs/2010ApJ...718...43P} {718, 43}
			
	\bibitem[\protect\citeauthoryear{{Poisson}, {D{\'e}moulin}, {L{\'o}pez Fuentes}
				\& {Mandrini}}{{Poisson} et~al.}{2016}]{Poisson2016_PropMagToung}
			{Poisson} M.,  {D{\'e}moulin} P.,  {L{\'o}pez Fuentes} M.,   {Mandrini} C.~H.,
			2016, \mn@doi [\solphys] {10.1007/s11207-016-0926-x}, \href
			{https://ui.adsabs.harvard.edu/abs/2016SoPh..291.1625P} {291, 1625}
			
	\bibitem[\protect\citeauthoryear{{Priest} \& {Forbes}}{{Priest} \&
				{Forbes}}{2002}]{priest2002}
			{Priest} E.~R.,  {Forbes} T.~G.,  2002, \mn@doi [\aapr]
			{10.1007/s001590100013}, \href
			{http://adsabs.harvard.edu/abs/2002A%26ARv..10..313P} {10, 313}
				
	\bibitem[\protect\citeauthoryear{{Romano}, {Pariat}, {Sicari}  \&
					{Zuccarello}}{{Romano} et~al.}{2011}]{Romano2011_SolEru_Trig_OppHel}
				{Romano} P.,  {Pariat} E.,  {Sicari} M.,   {Zuccarello} F.,  2011, \mn@doi
				[\aap] {10.1051/0004-6361/201014437}, \href
				{https://ui.adsabs.harvard.edu/abs/2011A&A...525A..13R} {525, A13}
				
	\bibitem[\protect\citeauthoryear{{Russell}, {Demoulin}, {Hornig}, {Pontin}  \&
					{Candelaresi}}{{Russell} et~al.}{2019}]{Russell2019}
				{Russell} A.~J.~B.,  {Demoulin} P.,  {Hornig} G.,  {Pontin} D.~I.,
				{Candelaresi} S.,  2019, \mn@doi [\apj] {10.3847/1538-4357/ab40b4}, \href
				{https://ui.adsabs.harvard.edu/abs/2019ApJ...884...55R} {884, 55}
				
	\bibitem[\protect\citeauthoryear{{Savcheva}, {Green}, {van Ballegooijen}  \&
					{DeLuca}}{{Savcheva} et~al.}{2012}]{savcheva2012a}
				{Savcheva} A.~S.,  {Green} L.~M.,  {van Ballegooijen} A.~A.,   {DeLuca} E.~E.,
				2012, \mn@doi [\apj] {10.1088/0004-637X/759/2/105}, \href
				{http://adsabs.harvard.edu/abs/2012ApJ...759..105S} {759, 105}
				
	\bibitem[\protect\citeauthoryear{{Schou}, {Scherrer}, {Bush}, {Wachter}  \& {et
						al}}{{Schou} et~al.}{2012}]{schou2012}
				{Schou} J.,  {Scherrer} P.~H.,  {Bush} R.~I.,  {Wachter} R.,   {et al} 2012,
				\mn@doi [\solphys] {10.1007/s11207-011-9842-2}, \href
				{http://adsabs.harvard.edu/abs/2012SoPh..275..229S} {275, 229}
				
	\bibitem[\protect\citeauthoryear{{Schuck}}{{Schuck}}{2008}]{schuck2008}
				{Schuck} P.~W.,  2008, \mn@doi [\apj] {10.1086/589434}, \href
				{http://adsabs.harvard.edu/abs/2008ApJ...683.1134S} {683, 1134}
		
	\bibitem[\protect\citeauthoryear{{Sun}, {Bobra}, {Hoeksema}  \& {et al}}{{Sun}
					et~al.}{2015}]{SunX2015_WhyFlareRichCMELess}
				{Sun} X.,  {Bobra} M.~G.,  {Hoeksema} J.~T.,   {et al} 2015, \mn@doi [\apjl]
				{10.1088/2041-8205/804/2/L28}, \href
				{https://ui.adsabs.harvard.edu/abs/2015ApJ...804L..28S} {804, L28}
				
	\bibitem[\protect\citeauthoryear{Taylor}{Taylor}{1974}]{Taylor1974}
				Taylor J.~B.,  1974, \mn@doi [Physical Review Letters]
				{10.1103/PhysRevLett.33.1139}, 33, 1139
				
	\bibitem[\protect\citeauthoryear{{Thalmann}, {Moraitis}, {Linan}, {Pariat},
					{Valori}  \& {Dalmasse}}{{Thalmann} et~al.}{2019}]{Thalmann2019_MagHel_ARs}
				{Thalmann} J.~K.,  {Moraitis} K.,  {Linan} L.,  {Pariat} E.,  {Valori} G.,
				{Dalmasse} K.,  2019, \mn@doi [\apj] {10.3847/1538-4357/ab4e15}, \href
				{https://ui.adsabs.harvard.edu/abs/2019ApJ...887...64T} {887, 64}
				
	\bibitem[\protect\citeauthoryear{{Tian} \& {Alexander}}{{Tian} \&
					{Alexander}}{2008}]{tian2008b}
				{Tian} L.,  {Alexander} D.,  2008, \mn@doi [\apj] {10.1086/524129}, \href
				{http://adsabs.harvard.edu/abs/2008ApJ...673..532T} {673, 532}
				
	\bibitem[\protect\citeauthoryear{{Tian}, {Alexander}  \& {Nightingale}}{{Tian}
					et~al.}{2008}]{tian2008a}
				{Tian} L.,  {Alexander} D.,   {Nightingale} R.,  2008, \mn@doi [\apj]
				{10.1086/589492}, \href {http://adsabs.harvard.edu/abs/2008ApJ...684..747T}
				{684, 747}
				
	\bibitem[\protect\citeauthoryear{{Valori}, {Pariat}, {Anfinogentov}  \& {et
						al}}{{Valori} et~al.}{2016}]{Valori2016}
				{Valori} G.,  {Pariat} E.,  {Anfinogentov} S.,   {et al} 2016, \mn@doi [\ssr]
				{10.1007/s11214-016-0299-3}, 201, 147
				
	\bibitem[\protect\citeauthoryear{{Vemareddy}}{{Vemareddy}}{2015}]{Vemareddy2015_HelEne}
				{Vemareddy} P.,  2015, \mn@doi [\apj] {10.1088/0004-637X/806/2/245}, \href
				{http://adsabs.harvard.edu/abs/2015ApJ...806..245V} {806, 245}
				
	\bibitem[\protect\citeauthoryear{{Vemareddy}}{{Vemareddy}}{2017}]{Vemareddy2017_homologus}
				{Vemareddy} P.,  2017, \mn@doi [\apj] {10.3847/1538-4357/aa7ff4}, \href
				{http://adsabs.harvard.edu/abs/2017ApJ...845...59V} {845, 59}
				
	\bibitem[\protect\citeauthoryear{{Vemareddy}}{{Vemareddy}}{2019a}]{Vemareddy2019_DegEle}
				{Vemareddy} P.,  2019a, \mn@doi [\mnras] {10.1093/mnras/stz1020}, \href
				{https://ui.adsabs.harvard.edu/abs/2019MNRAS.486.4936V} {486, 4936}
				
	\bibitem[\protect\citeauthoryear{{Vemareddy}}{{Vemareddy}}{2019b}]{vemareddy2019_VeryFast}
				{Vemareddy} P.,  2019b, \mn@doi [\apj] {10.3847/1538-4357/ab0200}, \href
				{https://ui.adsabs.harvard.edu/abs/2019ApJ...872..182V} {872, 182}
				
	\bibitem[\protect\citeauthoryear{{Vemareddy} \& {D{\'e}moulin}}{{Vemareddy} \&
					{D{\'e}moulin}}{2017}]{Vemareddy2017_SucHelInj}
				{Vemareddy} P.,  {D{\'e}moulin} P.,  2017, \mn@doi [\aap]
				{10.1051/0004-6361/201629282}, \href
				{https://ui.adsabs.harvard.edu/abs/2017A&A...597A.104V} {597, A104}
				
	\bibitem[\protect\citeauthoryear{{Vemareddy} \& {Wiegelmann}}{{Vemareddy} \&
					{Wiegelmann}}{2014}]{vemareddy2014_Quasi_Stat}
				{Vemareddy} P.,  {Wiegelmann} T.,  2014, \mn@doi [\apj]
				{10.1088/0004-637X/792/1/40}, \href
				{http://adsabs.harvard.edu/abs/2014ApJ...792...40V} {792, 40}
				
	\bibitem[\protect\citeauthoryear{{Vemareddy}, {Ambastha}  \&
					{Maurya}}{{Vemareddy} et~al.}{2012a}]{vemareddy2012_sunspot_rot}
				{Vemareddy} P.,  {Ambastha} A.,   {Maurya} R.~A.,  2012a, \mn@doi [\apj]
				{10.1088/0004-637X/761/1/60}, \href
				{http://adsabs.harvard.edu/abs/2012ApJ...761...60V} {761, 60}
				
	\bibitem[\protect\citeauthoryear{{Vemareddy}, {Ambastha}, {Maurya}  \&
					{Chae}}{{Vemareddy} et~al.}{2012b}]{vemareddy2012_hinj}
				{Vemareddy} P.,  {Ambastha} A.,  {Maurya} R.~A.,   {Chae} J.,  2012b, \mn@doi
				[\apj] {10.1088/0004-637X/761/2/86}, \href
				{http://adsabs.harvard.edu/abs/2012ApJ...761...86V} {761, 86}
				
	\bibitem[\protect\citeauthoryear{{Vourlidas}, {Lynch}, {Howard}  \&
					{Li}}{{Vourlidas} et~al.}{2013}]{Vourlidas2013}
				{Vourlidas} A.,  {Lynch} B.~J.,  {Howard} R.~A.,   {Li} Y.,  2013, \mn@doi
				[\solphys] {10.1007/s11207-012-0084-8}, \href
				{http://adsabs.harvard.edu/abs/2013SoPh..284..179V} {284, 179}
				
	\bibitem[\protect\citeauthoryear{Wheatland, Sturrock  \& Roumeliotis}{Wheatland
					et~al.}{2000}]{Wheatland2000}
				Wheatland M.~S.,  Sturrock P.~A.,   Roumeliotis G.,  2000, \mn@doi [\apj]
				{10.1086/309355}, 540, 1150
				
	\bibitem[\protect\citeauthoryear{{Wiegelmann}}{{Wiegelmann}}{2004}]{wiegelmann2004}
				{Wiegelmann} T.,  2004, \mn@doi [\solphys]
				{10.1023/B:SOLA.0000021799.39465.36}, \href
				{http://adsabs.harvard.edu/abs/2004SoPh..219...87W} {219, 87}
				
	\bibitem[\protect\citeauthoryear{{Wiegelmann} \& {Inhester}}{{Wiegelmann} \&
					{Inhester}}{2010}]{wiegelmann2010}
				{Wiegelmann} T.,  {Inhester} B.,  2010, \mn@doi [\aap]
				{10.1051/0004-6361/201014391}, \href
				{http://adsabs.harvard.edu/abs/2010A%26A...516A.107W} {516, A107}
					
	\bibitem[\protect\citeauthoryear{Yamamoto \& Sakurai}{Yamamoto \&
						Sakurai}{2009}]{Yamamoto2009}
					Yamamoto T.~T.,  Sakurai T.,  2009, \mn@doi [\apj]
					{10.1088/0004-637x/698/1/928}, 698, 928
					
	\bibitem[\protect\citeauthoryear{{Zhang} \& {Flyer}}{{Zhang} \&
						{Flyer}}{2008}]{Zhangmei2008_HelBound}
					{Zhang} M.,  {Flyer} N.,  2008, \mn@doi [\apj] {10.1086/589993}, \href
					{https://ui.adsabs.harvard.edu/abs/2008ApJ...683.1160Z} {683, 1160}
					
	\bibitem[\protect\citeauthoryear{{Zhang} \& {Low}}{{Zhang} \&
						{Low}}{2005}]{zhangmei2005}
					{Zhang} M.,  {Low} B.~C.,  2005, \mn@doi [\araa]
					{10.1146/annurev.astro.43.072103.150602}, \href
					{http://adsabs.harvard.edu/abs/2005ARA%26A..43..103Z} {43, 103}
						
	\bibitem[\protect\citeauthoryear{{Zhang}, {Flyer}  \& {Low}}{{Zhang}
							et~al.}{2006}]{Zhangmei2006_MagFld_confin}
						{Zhang} M.,  {Flyer} N.,   {Low} B.~C.,  2006, \mn@doi [\apj] {10.1086/503353},
						\href {https://ui.adsabs.harvard.edu/abs/2006ApJ...644..575Z} {644, 575}
						
	\bibitem[\protect\citeauthoryear{{Zuccarello}, {Pariat}, {Valori}  \&
							{Linan}}{{Zuccarello} et~al.}{2018}]{Zuccarello2018_Thres_MagHel}
						{Zuccarello} F.~P.,  {Pariat} E.,  {Valori} G.,   {Linan} L.,  2018, \mn@doi
						[\apj] {10.3847/1538-4357/aacdfc}, \href
						{https://ui.adsabs.harvard.edu/abs/2018ApJ...863...41Z} {863, 41}
		\makeatother
\end{thebibliography}

\bsp	% typesetting comment
\label{lastpage}

%%%%%%%%%%%%%%%%%%%

\end{document}